\documentclass[10pt]{amsart}
\usepackage{amsmath}
\usepackage{latexsym}
\usepackage{amsfonts}
\usepackage{amssymb}
\usepackage{amsthm}
\usepackage{bbm,dsfont}
\usepackage{graphicx}

\newtheorem{theorem}{Theorem}
\newtheorem{lemma}{Lemma}
\newtheorem{corollary}{Corollary}

\theoremstyle{definition}

\newcommand{\R}{\mathbb R} 
\newcommand{\C}{\mathbb C} 

\newcommand{\hi}{\mathcal{K}} 
\newcommand{\ip}[2]{\left(\,#1\,|\,#2\,\right)} 
\newcommand{\mip}[2]{\langle\,#1\,|\,#2\,\rangle} 
\newcommand{\tr}{\textrm{tr}} 
\newcommand{\id}{\mathbbm{1}} 
\newcommand{\nul}{\mathbbm{O}} 
\newcommand{\fii}{\varphi}

\newcommand{\ve}{\mathbf{e}} 
\newcommand{\vf}{\mathbf{f}} 
\newcommand{\vsigma}{\boldsymbol{\sigma}} 

\newcommand{\B}{\mathcal{B}}
\newcommand{\Hy}{\mathcal{H}}
\newcommand{\F}{\mathcal{F}}
\newcommand{\Ha}{\mathcal{H}}

\newcommand{\mL}{\mathcal{L}}
\newcommand{\mS}{\mathcal{S}}

\begin{document}

\title{Coexistence of Qubit Effects}

\author{Paul Busch}
\address{Department of Mathematics, University of York, York, UK}
\email{pb516@york.ac.uk}

\author{Heinz-J\"urgen Schmidt}
\address{Universit\"at Osnabr\"uck, Fachbereich Physik, Osnabr\"uck, Germany}
\email{hschmidt@uni-osnabrueck.de}

\begin{abstract}
Two quantum events, represented by positive operators (effects), are
{\em coexistent} if they can occur as possible outcomes in a single
measurement scheme. Equivalently, the corresponding effects are
coexistent if and only if they are contained in the ranges of a
single (joint) observable. Here we give several equivalent characterizations 
of coexistent pairs of qubit effects. We also establish the equivalence between
our results and those obtained independently by other authors. 
Our approach makes explicit use of the Minkowski space geometry inherent 
in the four-dimensional real vector space of selfadjoint operators in a 
two-dimensional complex Hilbert space.
\end{abstract}

\maketitle

\section{Introduction}\label{sec:introduction}

It is a fundamental result of the quantum theory of measurement that
pairs of observables represented by noncommuting selfadjoint
operators cannot be measured together. The joint measurability of
two observables $A,B$ entails that for every state (density
operator) $\rho$ there is a joint probability distribution of the
observables such that the probability of obtaining a value of $A$ in
a (Borel) subset $X$ of $\mathbb R$ and a value of $B$ in a subset $Y$
of $\R$  is given by $\tr[\rho G(X\times Y)]$, where $G(X\times Y)$
is a positive operator. The probabilities for $A$ and $B$ alone are
included as the {\em marginal} distributions $X\mapsto \tr[\rho
G(X\times\R)]$, $Y\mapsto\tr[\rho G(\R\times Y)]$, respectively. The
operators $G(X\times\R)$ and $G(\R\times Y)$ coincide with the
spectral projections $E^A(X)$ and $E^B(Y)$ of $A$ and $B$,
respectively. From this it follows that $A$ and $B$ commute and that
the operators $G(X\times Y)$ are the projection operators
$E^A(X)E^B(Y)$.

For observables $E,F$ represented as positive operator 
measures (POMs) (say with values in $\R$), the existence of a joint
observable does not in general require the commutativity of $E$ and
$F$. Observables $E,F$ are said to be jointly measurable if there exists a joint
observable $G$ (with values in $\R^2$) of which they are marginals.
The positive operators (effects) $E(X)$, $F(Y)$ in the ranges of $E$
and $F$ are then contained in the range of a single observable
($G$). A collection of effects are called {\em coexistent} if
they are contained in the range of a single POM.

The fact that not all pairs of observables are jointly measurable 
marks a fundamental distinction between quantum mechanics and classical mechanics. 
The extension of the notion of observables to include general POMs gives room for many
families of observables to be jointly measurable, and it 
becomes important to determine what price is to be paid for 
reconciling this classical feature with the underlying quantum structure.
Since noncommuting {\em sharp} observables (i.e. projection valued measures) 
are never jointly measurable it is clear that the joint measurability of noncommuting
observables requires these observables to be {\em unsharp} (i.e. POMs that 
are not projection valued). 

The impossibility of joint measurements of noncommuting sharp observables can be 
presented as a consequence of the {\em no-cloning theorem} \cite{Wer98}. In fact, 
if unknown states could be cloned, this could be utilized to send identical copies into 
measuring devices for two or more noncommuting observables, thus rendering 
simultaneously the distributions of values for the original system. The relationship 
between {\em approximate} quantum cloning and joint measurements has been 
a subject of subsequent investigations (see, e.g., \cite{DArSa00,DArMaSa01,BrAnBa06,FePa07}). 
It will be interesting to explore further connections between joint measurability and quantum 
information tasks. 

It is an open problem to give general, operationally significant conditions for
the joint measurability of two observables. The relationship between joint measurability and 
uncertainty relations has been studied in some depth in the past two decades and is reviewed 
in \cite{BuHeLa07} for the position-momentum case and in \cite{BuSh06} for qubit experiments 
in the specific manifestation of Mach-Zehnder interferometry. In these studies it has been shown 
that the relation of joint measurability is an important structural feature of the set of quantum 
observables that is intimately linked with other features, notably the degree of unsharpness 
of an observable and appropriate metric structures. 

The present paper is a contribution to the emerging programme of investigating the structure 
of the set of observables, which should complement current studies of the dual structure of the 
set of quantum states. We will address the special case of two {\em simple} observables (having 
just two possible values) for a qubit system (represented by a two-dimensional Hilbert space). 
In this simplest possible case the joint measurability of two simple observables is equivalent to 
the coexistence of a pair of effects. 

The special case of two qubit effects of trace equal to unity had been solved by one of the 
authors a number of years ago \cite{Busch86}. In this case a simple operational interpretation 
of the coexistence condition has been given \cite{BuSh03}: it can be 
cast in the form of a trade-off relation for the degrees of unsharpness of the two coexistent 
observables, required by their noncommutativity. The problem was revisited in \cite{BuHei08} 
in the context of  an outline theory of {\em approximate} joint measurements of noncommuting 
sharp qubit observables. A coexistence condition for a wider (though not fully general) class 
of pairs of qubit effects was found subsequently in \cite{Liu-etal07}.

Here we deduce necessary and sufficient conditions for the coexistence of two arbitrary effects 
of a qubit system. In fact we give various alternative, equivalent forms of such conditions 
which arise from different choices of bases in the space of selfadjoint operators.
Since an earlier version of the present paper was made available as arXiv:0802.4167v1, 
two other papers presented independently different formulations of criteria for the coexistence 
of qubit effects, using different approaches \cite{StReHe08,Liu-etal08}.  The first of these appeared 
on the same day as our result (note that the coauthor T.~Heinosaari of that paper is the same 
person as the coauthor T.~Heinonen of \cite{BuHeLa07}). The authors of \cite{Liu-etal08} 
proved equivalence between their result and that of \cite{StReHe08}, and provided partly numerical
evidence suggesting equivalence with our results. Here we have obtained the coexistence
condition in a form that will explicitly be shown to be equivalent with the condition of \cite{Liu-etal08}.
We believe that our approach, which is based on the order and convex structures of the set of effects,
lends itself best to generalizations to higher dimensions.

The notions of effects and their coexistence were introduced by G.~Ludwig in the 1960s
in his fundamental work on the axiomatic foundation of quantum mechanics \cite{DBPT}.
We dedicate this work to the memory of G\"unther Ludwig (1918--2007).

\section{Coexistent pairs of effects}\label{sec:simple}

Let $\hi$ be a complex Hilbert space with inner product $\ip{}{}$,
and let $\mL\equiv [\nul,\id]$ denote the set of effects, that is,
all operators $a$ such that $\nul\prec a\prec \id$. Here $\nul$ and
$\id$ represent the null and identity operators, respectively, and
$\prec$ denotes the usual ordering of selfadjoint operators: $a\prec
b$ (equivalently, $b\succ a$) if $\ip{\fii}{a\fii}\le\ip{\fii}{
b\fii}$ for all $\fii\in\hi$.

Any effect $e$ together with its complement effect $e'=\id-e$ forms
a simple observable. In general, an observable with finitely many
values is determined essentially by a set of effects
$\{a_1,a_2,\dots,a_n\}$, where the indices label the values, $a_k$
is the effect that determines the probabilities for the outcome
labeled with $k$, and $\sum_k a_k=\id$.

\begin{lemma}
Two effects $e,f$ are coexistent if and only if there are effects $a,b\in\mL$, such that
\begin{equation}\label{coex-ef}
a\prec e\prec b,\quad a\prec f\prec b, \quad a+b=e+f.
\end{equation}
\end{lemma}
\noindent{\em Proof.}
In fact, these inequalities are necessary and sufficient for each element of  the set of operators
\begin{equation}\label{joint-ef}
\{a,e-a,f-a,\id-e-f+a\}.
\end{equation}
to be effects. This set thus defines an observable whose range
contains the effects $e$ and $f$ as well as $e'$ and $f'$; hence it
constitutes a joint observable for the simple observables given by
$\{e,e'\}$ and $\{f,f'\}$. \qed

For later reference we note a few well-known results.
\begin{lemma} \label{prec-coex}
Effects $e,f\in\mL$ are coexistent if (a) or (b) hold:\\
(a) $e\prec f$ or $e\succ f$ $e\prec f'$ or $e\succ f'$;\\
(b) $[e,f]=\nul$.\\
In particular, $e,e'$ are coexistent.
\end{lemma}
\noindent {\em Proof.} (a) Let $e\prec f$. Take $a=e,b=f$, then
$\nul\prec e=a\prec e,f\prec e+f-e=f=b\prec\id$. If $e\prec f'$,
take $a=\nul,b=e+f$, then $\nul\prec e,f\prec e+f-\nul=b\prec \id$.
The other two cases are treated similarly.\\
(b) If $e,f$ commute then the the operators $ef,\ ef',\ e'f,\ e'f'$ are effects 
which add up to $\id$ and constitute a joint observable for $e,f$.\\
Finally, choose $f=e'$, then $e\prec f'=e$, so $e,e'$ are
coexistent.
 \qed
 
The cases (a) and (b) will be referred to as the {\em trivial} cases of coexistence.
We also note without proof that if at least one of two effects $e,f$
is a projection, then the effects are coexistent if and only if they
commute. In this case the joint observable (\ref{joint-ef}) is
uniquely determined by $e,f$ via $a=ef$.

\section{Geometric preliminaries}

\subsection{Minkowski space isomorphism}

In the case $\hi=\C^2$, selfadjoint operators are represented as
hermitian $2\times 2$ matrices. These form a 4-dimensional real
vector space $M_4$, spanned by the basis
\[
\sigma_0=\left(\begin{matrix} 1&0\\ 0&1 \end{matrix}\right),\quad
\sigma_1=\left(\begin{matrix} 0&1\\ 1&0 \end{matrix}\right),\quad
\sigma_2=\left(\begin{matrix} 0&-i\\ i&0 \end{matrix}\right),\quad
\sigma_3=\left(\begin{matrix} 1&0\\ 0&-1 \end{matrix}\right).
\]

For $x\in M_4$ we have $x\succ\nul$  exactly when the eigenvalues
of $x$ are non-negative. We note also that 
$x \succ\nul$ is equivalent to $\mip{x}{x}\ge 0$ und $x_0\ge 0$.

We define $x{\succ_{o}}\nul$ (equivalently
$\nul\prec_o x$ ) to mean that $x\succ\nul$ and at least one
eigenvalue of $x$ is equal to zero. Then for $x,y\in M_4$, $x\succ_o
y$ (or $y\prec_o x$) is defined to mean $x-y\succ_o\nul$.

Next we define the bilinear form
\[
\mip{x}{y}:=x_0y_0-\sum_{i=1}^3x_iy_i=x_0y_0-\mathbf{x}\cdot\mathbf{y},
\]
where $x=\sum_{i=0}^3x_i\sigma_i$, $y=\sum_{i=0}^3y_i\sigma_i$.

We note without proof the following fact.
\begin{theorem}
$(M_4,\mip{}{},\prec,\prec_o)$ is isomorphic to the 4-dimensional Minkowski space.
\end{theorem}
\noindent Accordingly, we will apply freely the terminology of
Minkowski geometry and refer to $\mip{}{}$ as the (Minkowski) scalar product. 
We use the same notation for vectors and for points in $M_4$ as an affine space.

The {\em forward} and {\em backward light cones} of an element $x\in M_4$ are 
defined as the sets
\begin{equation*}
\F(x):=\{y\in M_4:\, x\prec_o y\},\quad \B(x):=\{y\in M_4:\, x\succ_o y\}.
\end{equation*}
A vector $x\in M_4$ is called {\em lightlike} if $\mip{x}{x}=0$. If
$\mip{x}{x} >0$ or $<0$, the vector $x$ is called {\em timelike} or
{\em spacelike}, respectively. Then $x\prec_o y$ is equivalent to
$y-x$ being lightlike and $y_0-x_0\ge 0$. Elements $x,y\in M_4$ are
called {\em spacelike related}, $x\,\sigma\, y$, if $\mip{x-y}{x-y}<0$.

The set of effects can now be written as ($\mathrm{conv}(X)$ denotes the convex
hull of a set $X$)
\[
\mL=[\nul,\id]=\mathrm{conv}(\F(\nul))\cap\mathrm{conv}(\B(\id))
\]
$\mL$ is convex and compact, that is, it includes its boundary 
$(\F(\nul)\cup\B(\id))\cap\mL$.

The Minkowski scalar product $\mip{e}{f}$ admits a simple physical meaning
if $e$ and $f$ are effects: it is equal to the probability of joint occurrence 
$\ip{\Phi}{e\otimes f \,\Phi}$ if the effects $e$ and $f$ are measured by, say, Alice and Bob at
a two-particle system in the entangled (singlet) state
$\Phi=\frac{1}{\sqrt{2}}(\psi_+\otimes\psi_--\psi_-\otimes\psi_+)$.

\subsection{Properties of spacelike related effects $e,f$ in $M_4$}

\begin{figure}
\includegraphics[width=7cm]{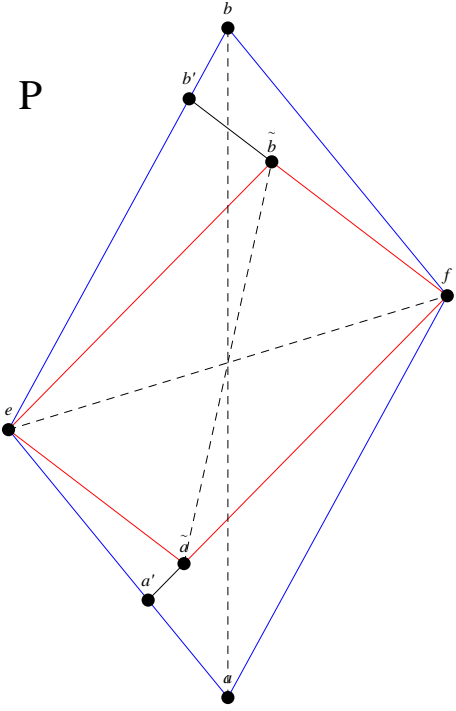}
\caption{ Illustration of the proof of Lemma \ref{ab-coex}.}\label{FIG_G1}
\end{figure}

\begin{lemma}\label{s-not-collin}
If effects $e,f\in\mL$ are spacelike related, $e\sigma f$, then the pairs $e,f$ and $e',f'$ 
are each linearly independent.
\end{lemma}
\noindent{\em Proof.} If $e,f$ are collinear so that (say) $f=\kappa e$ for some 
$\kappa \ge 0$, then $e-f=(1-\kappa)e$, and this is a timelike or lightlike vector. 
Similarly, if (say) $\id-f=\kappa(\id-e)$, then 
$e-f=-(1-\kappa)(\id-e)$ is timelike or lightlike. \qed

\begin{lemma}\label{ab-coex}
Let $e,f\in\mL$ be spacelike related ($e\,\sigma f$), and let $a,b\in\mL$ be
such that $a\prec e\prec b$, $a\prec f\prec b$ and $a+b=e+f$. Then
there exist $\tilde{a},\tilde{b}\in\mL$ such that $\tilde{a}\prec_o
e\prec_o \tilde{b}$, $\tilde{a}\prec_o f\prec_o \tilde{b}$ and
$\tilde{a}+\tilde{b}=e+f$.
\end{lemma}
\noindent{\em Proof.} Let $P$ be the 2-dimensional plane containing
$e,f,a$ and hence $b$ (see Fig.~\ref{FIG_G1}). In $P$ the forward and
backward light cones degenerate to lines. Since $e\,\sigma f$, the
forward and backward cones of $e$ and $f$ intersect in exactly one
point, respectively. Hence we define $\tilde a$ and $\tilde b$ by
$\B(e)\cap\B(f)\cap P=\{\tilde a\}$, $\F(e)\cap\F(f)\cap P=\{\tilde b\}$. The
lines $\ell(e,\tilde a)$ and $\ell(f,\tilde b)$ are parallel,
likewise $\ell(e,\tilde b)$ and $\ell(f,\tilde a)$. Hence $e,\tilde
b,f,\tilde a$ form the vertices of a parallelogram and $\tilde
a+\tilde b=e+f$. Due to the convexity of $\mL$, the element $\tilde
a\in\mL$ since the intersection of $\ell(\tilde a,f)$ and the line
segment $s(e,a)\subseteq\mL$ contains one element $a'\in\mL$ and
$\tilde a$ is in the line segment $s(a',f)\subseteq\mL$. Analogously
it is shown that $\tilde b\in\mL$. \qed

\begin{lemma}\label{lem:commut}
Effects $e,f$ which are  
not collinear
are mutually commuting 
if and only if $\id$ lies in the subspace spanned by $e,f$. 
\end{lemma}
\noindent{\em Proof.}  If $e=e_0\id+\mathbf{e}\cdot \vsigma$ and
$f=f_0\id+\mathbf{f}\cdot\vsigma$ commute, then 
\begin{equation*}
0=[e,f]=[\mathbf{e}\cdot\vsigma,\mathbf{f}\cdot\vsigma]= 2 i (\mathbf{e}\times \mathbf{f})\cdot \vsigma
\end{equation*}
hence $\mathbf{e}$ and
$\mathbf{f}$ are collinear; since $e,f$ are  
not collinear, it
follows that $\id$ is in the span
of $e$ and $f$. Conversely, if $\id=xe+yf$ then $x\mathbf{e}+y\mathbf{f}=\nul$, so $[e,f]=\nul$.
\qed

For two effects $e,f\in M_4$ we define $M(e,f,\id)$ as the  Minkowski subspace of $M_4$ spanned
by $e,f$ and $\id$ and equipped with the orderings $\prec,\prec_o$
inherited from $M_4$. Note that if $e\sigma f$, then 
$M(e,f,\id)$ is 2-dimensional exactly when $e,f$ commute (Lemma \ref{lem:commut})
and otherwise 3-dimensional.

\begin{lemma}\label{monotone}
Let $T\subset M_4$ be a $3$-dimensional timelike subspace, i.~e.~a subspace containing at least one timelike vector,
such that $\id\in T$.
Then its $\mip{}{}$-orthogonal complement $T^\perp$ will be a one-dimensional spacelike subspace and
$M_4=T\oplus T^\perp$. The $\mip{}{}$-orthogonal linear projection $\pi:M_4\to T$ will be monotone,
i.e., if $a,b\in M_4$ and $a\prec b$, then $\pi(a)\prec\pi(b)$, and $\mip{}{}$-selfadjoint.
\end{lemma}
\noindent{\em Proof.} Each vector $b\in M_4$ can be uniquely written as $b=\pi(b)+b_\perp$,
such that $\pi(b)\in T$ and $b_\perp\in T^\perp$. Let $a,b\in M_4$. Then
$\mip{a}{\pi(b)}=\mip{\pi(a)+a_\perp}{\pi(b)}=\mip{\pi(a)}{\pi(b)}=\mip{\pi(a)}{\pi(b)+b_\perp}=\mip{\pi(a)}{b}$.
This proves $\pi$ being selfadjoint.\\
Concerning monotonicity it suffices to consider the case
$\nul=a\prec b$, i.~e.~$\mip{b}{b}\ge 0$ and $b_0\ge 0$, since $\pi$ is linear.
It follows that $\mip{b_\perp}{b_\perp}\le 0$ since $T^\perp$ is spacelike, and further
$0\le \mip{b}{b}=\mip{\pi(b)}{\pi(b)}+\mip{b_\perp}{b_\perp}\le\mip{\pi(b)}{\pi(b)}$. Moreover,
$0\le b_0=\mip{b}{\id}=\mip{b}{\pi(\id)}=\mip{\pi(b)}{\id}=\pi(b)_0$, using $\pi$ being selfadjoint.
Both inequalities together imply that $\nul\prec\pi(b)$, which concludes the proof.   \qed

\subsection{The Minkowski subspace $M_3$}

We will make use of a 3-dimensional Minkowski subspace
$M_3(\cong\R^3)$ of $M_4$, defined as the linear span of
$\sigma_0,\sigma_1,\sigma_2$, with the orderings $\prec,\prec_0$
carried over from $M_4$. For $x,y\in M_3$, define
\begin{equation*}
x{\times_o} y:=\left(\begin{matrix}x_0\\ x_1\\ x_2\end{matrix}\right)\times_o
\left(\begin{matrix}y_0\\ y_1\\ y_2\end{matrix}\right):=
\left(\begin{matrix}x_1y_2-x_2y_1\\ x_0y_2-x_2y_0\\ x_1y_0-x_0y_1\end{matrix}\right).
\end{equation*}
Hence $x\times_o y$ is the usual vector product, but with spacelike
components inverted. We will use freely the following properties.
\begin{lemma}\label{lem:rules}
Let $x,y\in M_3$. Then
\begin{eqnarray*}
x\times_o y&=&-y\times_o x;\\
x\times_o(y\times_o z)&=&y\mip{x}{z}-z\mip{x}{y};\\
\mip{x}{x\times_o y}&=&\mip{y}{x\times_o y}=0;\\
\mip{x\times_o y}{\tilde x\times_o \tilde y}&=&
\mip{x}{\tilde x}\mip{y}{\tilde y}-\mip{x}{\tilde y}\mip{\tilde x}{y}.
\end{eqnarray*}
Furthermore, $x\times_o y=\nul$ if and only if  $x,\ y$\ are collinear.
\end{lemma}

We note that the subspace $M(e,f,\id)$ can be
identified with (a subspace of)  $M_3$ since $e,f$ can be unitarily transformed into
elements of $M_3$.

We now introduce three basis systems in $M_3$ and give some properties that are useful in what follows.

For $e,f\in \mL\cap M_3$, we define two vectors
\begin{equation}\label{eqn:gpm}
g\equiv e\times_o f,\quad g'\equiv e'\times_o f'.
\end{equation}
By definition they are in the subspace $\mip{}{}$-perpendicular to the vector
\begin{equation}\label{eqn:d=e-f}
d\equiv e-f=f'-e'\;,
\end{equation}
so that the triple $\{g,g',d\}$ forms a basis of $M_3$ if $g,g'$ are linearly independent.
Hence,
\begin{equation}\label{eqn:gpm-perp-d}
\mip{g}{d}=\mip{g'}{d}=0.
\end{equation}

\begin{lemma}\label{lem:g+g-}
For $e,f\in\mL\cap M_3$, the following statements hold:
\begin{itemize}
\item[(a)] The vectors $g$ and $g'$ are both nonzero iff neither $e,f$ nor $e',f'$ are collinear.
In particular, $g\ne\nul\ne g'$ if $e\sigma f$.
\item[(b)] The vectors $g$, and $g'$ are both spacelike vectors whenever they are nonzero.
\item[(c)]  If $e\ne f$, the vectors $g,\ g'$ are linearly independent iff 
$[e,f]\ne \nul$.
\item[(d)] $g\times_og'$ is spacelike iff $e\sigma f$ and $[e,f]\ne\nul$. 
\end{itemize}
\end{lemma}
\noindent{\em Proof.} 
(a) If $e\sigma f$, then by Lemma \ref{s-not-collin} the vector pairs $e,f $ and $e',f'$ are both linearly 
independent, and thus, by Lemma \ref{lem:rules}, $g=e\times_of$ and $g'=e'\times_of'$ are nonzero.\\
(b) Two timelike or lightlike vectors are never $\mip{}{}$-perpendicular unless they are collinear and lightlike.
Thus, since $e$ is timelike or lightlike, then $g$ must be spacelike. A similar argument applies to $g'$.\\
(c) Next we note that
\[
\mip{e}{e'\times_o f'}-\mip{f}{e'\times_o f'}=\mip{d}{e'\times_o f'}=0,
\]
and use this to compute:
\begin{eqnarray*}
g\times_o g'&=& (e\times_o f)\times_o (e'\times_o f')
= -e\mip{f}{e'\times_o f'}+f\mip{e}{e'\times_o f'}\\
&=& -d\mip{e}{e'\times_o f'}
= -d\mip{e}{\id\times_o\id +\id\times_o d+e\times_o f}\\
&=& -d\mip{e}{\id\times_o d}=-d\mip{\id}{e\times f},
\end{eqnarray*}
so
\begin{equation}\label{g+xg-}
g\times_o g'= -d(e_1f_2-e_2f_1)=\pm d\,|\mathbf{e}\times\mathbf{f}|.
\end{equation}
Thus if $d\ne\nul$, then $g\times_og'\ne\nul$ exactly when $e,f$ do not commute.\\ 
(d) The last statement follows equally immediately by inspection of (\ref{g+xg-}).
 \qed

We compute the inner products:
\begin{eqnarray}
C(e,f)&\equiv&\mip{g}{g}=
\mip{e}{e}\mip{f}{f}-\mip{e}{f}^2\,;\label{cef}\\
C(e',f')&\equiv&\mip{g'}{g'}=
\mip{e'}{e'}\mip{f'}{f'}-\mip{e'}{f'}^2\,;\label{cef'}\\
D(e,f)&\equiv& D(e',f')\equiv\mip{g}{g'}\nonumber\\
&=& \mip{e}{e'}\mip{f}{f'}-\mip{e}{f'}\mip{e'}{f}\,;\label{def}\\
\Delta(e,f)&\equiv&\Delta(e',f')\equiv\mip{g\times_o g'}{g\times_o g'}\nonumber\\
&=&\mip{g}{g}\mip{g'}{g'}-\mip{g}{g'}^2
=\mip{d}{d}|\ve\times\vf|^2\;.\label{delta}
\end{eqnarray}
The first two quantities are non-positive since $e,f,e',f'$ are timelike or lightlike. 
(One can also directly use the fact that $e\times_o f$ and $e'\times_o f'$ 
are spacelike whenever they are nonzero.)
We will also show that the third term $\mip{g}{g'}>0$ if $e\sigma f$. 
Furthermore, as is seen as a direct consequence of eq.~(\ref{g+xg-}), the last term is negative
if and only if $e\sigma f$ and $[e,f]\ne\nul$; given
the above explicit expressions, this means  that the spacelike vectors $g,g'$ satisfy an 
inverted Schwarz inequality.

\begin{lemma}\label{detpos}
For all $e,f\in\mL$ 
 the following inequalities hold:
\begin{eqnarray}
C(e,f)&=&\mip{e}{e}\mip{f}{f}-\mip{e}{f}^2\le 0\,;\label{C<0}\\
C(e',f')&=&\mip{e'}{e'}\mip{f'}{f'}-\mip{e'}{f'}^2\le 0\,;\label{C'<0}\\
 D(e,f)&=&\mip{e}{e'}\mip{f}{f'}-\mip{e}{f'}\mip{f}{e'}> 0\quad\text{if\ }e\sigma f\,;\label{D>0}\\
 \Delta(e,f)&=&C(e,f)C(e',f')-D(e,f)^2< 0\quad\text{iff\ }e\sigma f \ \text{and\ }[e,f]\ne \nul.\label{Delta<0}
 \end{eqnarray}
\end{lemma}
\noindent{\em Proof.}
For the purposes of the proof we consider $M(e,f,\id)$ as embedded in $M_3$.
It remains to verify $D(e,f)>0$.
We show that $D(e,f)$ can be expressed as
\begin{equation}\label{D-newform}
D(e,f)=\tfrac 12\mip{e'+f'}{d\times_og}.
\end{equation}
In fact, using the rules for $\times_o$ and using $d=f'-e'$ we find: $\mip{e'+f'}{d\times_og}=
\mip{(e'+f')\times_od}{g}=2\mip{g'}{g}$. Now, $e\sigma f$, and so the forward-oriented timelike or 
lightlike vectors $e',f'$ are not collinear, so that  $e'+f'$ is timelike. Likewise, 
$d\times_og$ is timelike since this vector is $\mip{}{}$-perpendicular (in $M_3$) to two spacelike 
vectors. We show that $d\times_og$ is forward-oriented. This entails that the inner product of $e'+f'$ 
and $d\times_og$ is positive.\\
Thus we have to show that $\mip{\id}{d\times_og}>0$. First note that the vector $\sqrt{f_0}e-\sqrt{e_0}f$
is spacelike. In fact, otherwise one would have (say) $\sqrt{f_0}e\succ\sqrt{e_0}f$, so that $e_0\ge f_0$
and finally $e\succ\sqrt{e_0/f_0}f\succ f$, which contradicts $e\sigma f$. Now we have:
\begin{equation*}\begin{split}
0&\le -\mip{\sqrt{f_0}e-\sqrt{e_0}f}{\sqrt{f_0}e-\sqrt{e_0}f}
=2\sqrt{e_0f_0}\mip{e}{f}-f_0\mip{e}{e}-e_0\mip{f}{f}\\
&\le(e_0+f_0)\mip{e}{f}-f_0\mip{e}{e}-e_0\mip{f}{f}=\mip{\id}{e}\mip{e-f}{f}-\mip{\id}{f}\mip{e-f}{e}\\
&=\mip{\id}{d\times_o(e\times_of}=\mip{\id}{d\times_og}\,.\hspace{6cm}\qed
\end{split}
\end{equation*}

The vector pair $g,\ g'$ was found to be collinear if $e$ and $f$ commute. To remove this degeneracy, we 
also consider the basis $\{d,\ g,\ d\times g\}$ of mutually $\mip{}{}$-perpendicular vectors in $M_3$. 
We note the following for later use:
\begin{eqnarray}
\mip{e+f}{g}&=&\mip{e'+f'}{g'}=0\,;\label{efg}\\
\mip{e'+f'}{g}&=&\mip{e+f}{g'}=2\mip{\id}{g}=2(e_1f_2-e_2f_1)\nonumber\\
&=&2\,\text{sign}\,(e_1f_2-e_2f_1)\,|\ve\times\vf|\,;\label{gpm}\\
\mip{e+f}{d\times_og}&=&-2\mip{g}{g}=-2C(e,f)\,;\label{efdg}\\
\mip{e'+f'}{d\times_og}&=&2D(e,f)\,;\label{e'f'dg}\\
\mip{d\times_og}{d\times_og}&=&\mip{d}{d}\mip{g}{g}-\mip{d}{g}^2=\mip{d}{d}\;C(e,f)\,.\label{dgdg}
\end{eqnarray}

Using these identities is not hard to verify that
\begin{equation*}
g'=\frac{D(e,f)}{C(e,f)}\;g\ -\ \frac{e_1f_2-e_2f_1}{C(e,f)}\;d\times_og.
\end{equation*}
This confirms that $g$ and $g'$ are collinear exactly when the second term vanishes, that is, when $[e,f]=\nul$.

Finally we introduce a basis $\{d,h_+,h_-\}$ of $M_3$ where $h_+,h_-$ are distinct lightlike vectors 
orthogonal to $d$. Note that this presupposes that $e\sigma f$, so that $\mip{d}{d}<0$. 
We write $h_\pm=x_\pm g+y_\pm d\times_og$ and compute:
\begin{equation*}
0=\mip{h_\pm}{h_\pm}=x_\pm^2\mip{g}{g}+y_\pm^2\mip{d\times_og}{d\times_og}
=C(e,f)\left[x_\pm^2+y_\pm^2 \mip{d}{d}\right].
\end{equation*}
Here we have used the identity (\ref{dgdg}).
Thus we find (using a particular choice of the overall constant factor):
\begin{equation}\label{eqn:hpm}
h_\pm=\pm \sqrt{|\mip{d}{d}|}\;g+d\times_og\,.
\end{equation}

We compute:
\begin{eqnarray}
\mip{h_+}{h_-}&=&2C(e,f)\mip{d}{d}>0\,;\label{eqn:h+h-}\\
\mip{e+f}{h_\pm}&=&-2C(e,f)>0\,;\label{eqn:efhpm}\\
\mip{e'+f'}{h_\pm}&=&2\left[D(e,f)\pm\sqrt{|\mip{d}{d}|}\,(e_1f_2-e_2f_1)\right]\label{e'f'hpm}\\
&=&2\left[D(e,f)\pm \sqrt{|\mip{d}{d}|} |\ve\times\vf|\text{sign}\,(e_1f_2-e_2f_1)\,\right]\nonumber\\
&=&2\left[D(e,f)\pm\sqrt{|\Delta(e,f)|}\,\text{sign}(e_1f_2-e_2f_1)\,\right]>0\,.\nonumber
\end{eqnarray}
The last relation follows by application of the identity (\ref{gpm}).
These quantities are all positive in the present case of $e\sigma f$. 

Henceforth we will assume that $\text{sign}\,(e_1f_2-e_2f_1)\,=+1$ so that we can always replace
$e_1f_2-e_2f_1(>0)$ with $|\ve\times\vf|$. This can always be arranged by swapping $e$ and $f$ 
if necessary. Our main results in the next section will be given in a form that is invariant under this 
exchange.

\section{Coexistent pairs of qubit effects}\label{sec:qubit}

Next we consider the question of the coexistence of qubit effects $e,f$
which are spacelike related and not necessarily mutually commuting. 
The conditions obtained will ultimately be phrased in such a way that
they hold also in the trivial cases of coexistence.

\subsection{Reduction to $M_3$}

We first show that the coexistence of a pair of effects $e,f\in M_4$ can be studied
within $M_3$ (taking into account that the relation of coexistence is invariant under 
unitary transformations). The resulting conditions will be written in a form that is 
invariant under spatial rotation, using identities such as 
\[
(e_1 f_2-e_2 f_1)^2 =|\mathbf{e}\times\mathbf{f}|^2=-\frac{1}{12}\tr([e,f]^2)= \frac 14 \|[e,f]\|^2\,.
\]
In this way the result will be generally valid in $M_4$; the reduction to $M_3$
is only made for the sake of simplifying the proofs.

\begin{theorem}
 If $e,f\in\mL$ are
coexistent, then they are also coexistent in $M(e,f,\id)$, that is,
there is an effect $a\in\mL\cap M(e,f,\id)$ such that
\[
\nul\prec a\prec e,f\prec e+f-a\prec\id.
\]
\end{theorem}
\noindent{\em Proof.} If $e\prec f$ or $f\prec e$ or $[e,f]=\nul$ the claim follows directly. 
Hence we may consider the case where $e\sigma f$ and $M(e,f,\id)$ is a $3$-dimensional 
timelike subspace of $M_4$  and $\pi:M_4\to M(e,f,\id)$ the corresponding linear projection 
which, by Lemma \ref{monotone}, is monotone. Let $a\in M_4$, $a\prec e,f$ and 
$e+f-a\prec\id$. It follows that
\begin{eqnarray*}
\pi(a)&\prec&\pi(e)=e,\\
\pi(a)&\prec&\pi(f)=f,\\
\pi(e)+\pi(f)-\pi(a)=\pi(e+f-a)&\prec&\pi(\id)=\id.
\end{eqnarray*}
Hence $e,f$ are coexistent in $M(e,f,\id)$. \qed

An obvious corollary is that if effects $e,f\in M_4$ are coexistent and  
$M(e,f,\id)\subseteq M_3$, then $e,f$ are also coexistent as elements of $M_3$, and vice versa.

\subsection{Characterization of coexistence in $M_3$}\label{subsec:coex-M3}

We will use the same notation as in $M_4$ for the forward and backward cone of an element $x$ in $M_3$, 
namely, $\F(x)$, $\B(x)$.  The coexistence criterion of Lemma \ref{ab-coex} then states that effects $e,f$ 
in $M_3$ are coexistent if and only if there is an effect $a\in\B(e)\cap\B(f)$ such that $b=e+f-a$ is also an effect.
This is trivially satisfied if $e\prec f$ or $e\succ f$, for then one can choose $a=e, b=f$ in the first case and $a=f,b=e$ in the second. In these trivial cases one of the backward cones encloses the other, and they are disjoint unless $e,f$ are lightlike related. The case $e\sigma f$ is less trivial. We recall the following familiar fact.

\begin{lemma}\label{lem:hyp-BeBf}
Let $e,f\in\mL\cap M_3$ be spacelike related effects, $e\sigma f$.
Let $H$ be the plane passing through $\frac 12(e+f)$ which is
$\mip{}{}$-perpendicular to $d$. Then the
intersections $\Ha_a\equiv\B(e)\cap\B(f)$ and $\Ha_b\equiv\F(e)\cap\F(f)$ are the two
branches of a hyperbola $\Ha$ lying in $H$. 
\end{lemma}
\noindent{\em Proof.} Each of the conditions
$a\in\B(e)\cap\B(f)$ and $b=e+f-a\in\F(e)\cap\F(f)$ are equivalent to
$\mip{e-a}{e-a}=0=\mip{f-a}{f-a}$ and this gives
$\mip{\frac 12(e+f)-a}{e-f}=0$. Hence this intersection of the two cones lies actually in the plane $H$ and thus
is a conic section.
\qed

Let $e,f\in M_3$, with $e\,\sigma f$. 
 Writing $a=\frac 12(e+f)-v$,
the coexistence condition is now spelled out as follows:\\
(i) $a\in H$ is equivalent to
\begin{equation}\label{v-d}
 \mip{v}{d}=0.
 \end{equation}
 \noindent (ii) $a\in\Ha_a=\B(e)\cap\B(f)$ and $b=e+f-a\in\Ha_b=\F(e)\cap\F(f)$ are both equivalent to
\begin{eqnarray}
\mip{v\pm\tfrac 12d}{v\pm\tfrac 12d}&=&
\mip{v}{v}+\tfrac 14\mip{d}{d}=0,\label{hyp1}\\
 \ v_0&\ge& \tfrac 12|d_0| .\label{aBbF}
\end{eqnarray}
(Note that here we have utilized (i). We also remark that (\ref{aBbF}) can be replaced by the weaker
$v_0>0$. The sharper bound arises from the fact that $a\prec e,f$ implies 
$a_0=\frac 12(e_0+f_0)-v_0\le e_0,\,f_0$.)\\
\noindent (iii) The  conditions $a\succ \nul$ and $b=e+f-a\prec\id$
specify two bounded segments 
\begin{equation}\label{eqn:sasb}
\mS_a\equiv \B(e)\cap\B(f)\cap\mathrm{conv}(\F(\nul)),\quad 
\mS_b\equiv\B(e)\cap\B(f)\cap\mathrm{conv}(\F(e-f'))
\end{equation}
of admissible elements $a$ on the hyperbola branch $\Ha_a=\B(e)\cap\B(f)$.

Note that $\mS_a\neq\emptyset$ since $\nul\prec e,\,f$, so that $\B(e)\cap\B(f)$ cannot fall entirely
outside $\mL$.  Similarly, $\mS_b\neq\emptyset$ since $e,\,f\prec\id$, so that $\F(e)\cap\F(f)$ cannot fall entirely outside of $\mL$.
But it may happen that $\mS_a$ as well as $\mS_b$ degenerate into a single point. The 
coexistence conditions can thus be characterized geometrically.
\begin{lemma}
Let  $e,f\in\mL\cap M_3$, $e\sigma f$.  Then
\[
 e,f {\text \ coexistent } \Longleftrightarrow \mS_a\cap\mS_b\ne\emptyset\,.
\]
\end{lemma}
Since $e,f$ are coexistent if $[e,f]=\nul$, it follows that $\mS_a\cap\mS_b\ne\emptyset$ in the
commutative case. The following confirms further  trivial cases of coexistence:
\begin{eqnarray}
e\prec f' &\Longleftrightarrow& \mathrm{conv}(\F(\nul))\subseteq 
\mathrm{conv}(\F(e-f'))\Longrightarrow \mS_a\subseteq\mS_b\, ,\label{e-prec-f'}\\
e\succ f' &\Longleftrightarrow& \mathrm{conv}( \F(e-f'))\subseteq 
\mathrm{conv}(\F(\nul))\Longrightarrow \mS_b\subseteq\mS_a\, .\label{f'-prec-e}
\end{eqnarray}
This means conversely that
\begin{equation}
\mS_a\not\subseteq\mS_b\ \text{and}\ \mS_b\not\subseteq\mS_a\ \Longrightarrow\ e\sigma f'\,.
\end{equation}
The remaining trivial cases of coexistence, $e\prec f$ or $e\succ f$, cannot be characterized in terms of 
$\mS_a$, $\mS_b$ since the hyperbola and hence these sets are only defined if $e\sigma f$.

We proceed to find necessary and sufficient conditions for $\mS_a\cap\mS_b\ne\emptyset$ to be true.
The end points of the segments $\mS_a$, $\mS_b$ are determined by $a\succ_o\nul$, that is,
\begin{eqnarray}
0&=&\mip{a}{a}=\tfrac 14\mip{e+f}{e+f}+\mip{v}{v}-\mip{e+f}{v}\nonumber\\
&=&\mip{e}{f}-\mip{e+f}{v}\label{vvdd1}\quad\text{ for }\mS_a, \quad\text{[using (\ref{hyp1})]}\\ \label{vvdd12}
0&\le& a_0=\tfrac 12(e_0+f_0)-v_0,
\end{eqnarray}
and $e+f-a\prec_o\id$, that is,
\begin{eqnarray}
0&=&\mip{\id+a-e-f}{\id+a-e-f}\nonumber\\
&=& 1+\tfrac 14\mip{e+f}{e+f}+\mip{v}{v}-\mip{\id}{e+f}\nonumber\\
&&\qquad -2\mip{\id}{v}+\mip{e+f}{v}\nonumber\\
&=&\mip{e'}{f'}-\mip{e'+f'}{v}\quad\text{ for }\mS_b,\quad\text{[using (\ref{hyp1})]}\label{vvdd2}\\ \label{vvdd22}
0&\le& 1-e_0-f_0+a_0=\tfrac 12(e'_o+f'_0)-v_0.
\end{eqnarray}
Note that in (\ref{vvdd1}) and (\ref{vvdd2}) we have used (\ref{hyp1}).

Inequalities (\ref{vvdd12}) and (\ref{vvdd22}) already
follow from the remaining conditions and hence can be neglected as
far as equivalence to the coexistence of $e$ and $f$ is concerned. In fact,
geometrically, (\ref{vvdd12}) holds since the hyperbola $\B(e)\cap\B(f)$ intersects
$\F(\nul)$ in exactly two points, or touches $\F(\nul)$ at a single point
in special cases. On the other hand, $\B(e)\cap\B(f)\cap \B(\nul)=\emptyset$
or, in special cases, $\B(e)\cap\B(f)\cap \B(\nul)=\{ \nul \}$. Analogous arguments
apply to (\ref{vvdd22}).

\subsection{Main result}

We now deduce a set of coexistence conditions using the lightlike vectors
$h_\pm$  (eq.~(\ref{eqn:hpm})) for the parametrization of the plane
$H$ containing $\Ha$. The vectors $h_\pm$
will be found to determine the directions of the asymptotes of the hyperbola $\Ha$, 
which intersect in the point $\frac 12 (e+f)$.

We start with the hyperbola condition (\ref{hyp1}) and the linear equations 
(\ref{vvdd1}) and  (\ref{vvdd2}) which specify the segments $\mS_a$ and $\mS_b$, 
respectively. The end points of $\mS_a,\mS_b$ will be determined using the parametrization
$a=\frac 12(e+f)-\lambda h_+-\mu h_-$.\footnote{We will  use the same notation
$\mS_a,\mS_b$ for the representations of the segments in the $\lambda-\mu-$plane.} 

The equation (\ref{hyp1}) of the hyperbola now becomes (using (\ref{eqn:h+h-})):
\begin{equation}
\lambda\mu=-\frac 18\,\frac{\mip{d}{d}}{\mip{h_+}{h_-}}=\frac 1{16|C(e,f)|}\,.
\end{equation}
Note that now the asymptotes of the hyperbola in the $\lambda$-$\mu$-plane are perpendicular,  and $\mu(\lambda)$ is 
a monotonic function. If $\lambda\to\infty$, then $\mu\to 0$, and the vector $v=\lambda h_++\mu h_-$ pointing from $\frac 
12(e+f)$ to a point on the hyperbola (in $M_3$) approaches $\lambda h_+$, which is thus seen to be in the direction of 
an asymptote. Similarly, the direction of the other asymptote is given by $h_-$.

\begin{figure}
\includegraphics[width=10cm]{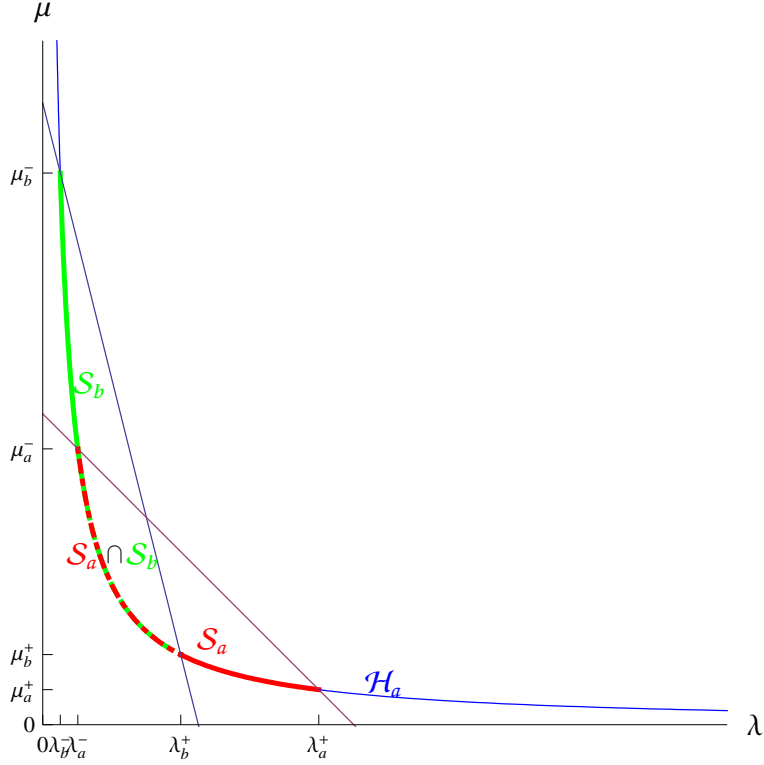}
\caption{\label{FIG_G5}Representation of the hyperbola branch $\B(e)\cap\B(f)$ in the 
$\lambda-\mu-$plane, with an indication of the coordinates of the end points of the segments
$\mS_a,\mS_b$.}
\end{figure}

The equations defining the line segments $\mS_a,\mS_b$ are:
\begin{eqnarray}
 \lambda\mip{e+f}{h_+}+\mu\mip{e+f}{h_-}&=&\mip{e}{f}\quad\text{for\ }\mS_a\, ,\\
 \lambda\mip{e'+f'}{h_+}+\mu\mip{e'+f'}{h_-}&=&\mip{e'}{f'}\quad\text{for\ }\mS_b\, .
\end{eqnarray}
These represent straight lines with negative gradients and intersecting the hyperbola in the first quadrant, see Fig.~\ref{FIG_G5}.
Using the expressions (\ref{eqn:efhpm}), (\ref{e'f'hpm}) for the coefficients, these linear equations can be
rewritten as:
\begin{eqnarray}
 \mu&=&\frac{\mip{e}{f}}{2|C(e,f)|}-\lambda\quad\text{for\ }\mS_a\, ,\\
 \mu&=&\frac{\mip{e'}{f'}}{2(D(e,f)-\sqrt{|\Delta(e,f)|})}-\lambda\frac{D(e,f)+\sqrt{|\Delta(e,f)|}}{D(e,f)-\sqrt{|\Delta(e,f)|}}\quad\text{for\ }\mS_b\, .
\end{eqnarray}
The first line has a fixed negative slope $-1$ and the second is always steeper downward.
We denote the intersection points of these lines with the hyperbola 
$(\lambda_{a}^\pm,\,\mu_{a}^\pm)$ and $(\lambda_{b}^{\pm},\mu_{b}^{\pm})$. 
Using the expressions (\ref{eqn:efhpm}), (\ref{e'f'hpm}) for the coefficients, we find:
\begin{equation*}\begin{split}
\lambda_{a}^{\pm}&=\frac{1}{2\mip{e+f}{h_+}}\left\{\mip{e}{f}\pm\sqrt{\mip{e}{f}^2-
\frac{\mip{e+f}{h_+}\mip{e+f}{h_-}}{4|C(e,f)|}}\right\}\nonumber\\
&=\frac 1{4|C(e,f)|}\left\{\mip{e}{f}\pm\sqrt{\mip{e}{e}\mip{f}{f}} \right\}
>0\,;\\
\lambda_{b}^{\pm}&=\frac{1}{2\mip{e'+f'}{h_+}}\left\{\mip{e'}{f'}\pm\sqrt{\mip{e'}{f'}^2-
\frac{\mip{e'+f'}{h_+}\mip{e'+f'}{h_-}}{4|C(e,f)|}}\right\}\nonumber\\
&=\frac 1{4(D(e,f)+\sqrt{|\Delta(e,f)|})}\left\{\mip{e'}{f'}\pm\sqrt{\mip{e'}{e'}\mip{f'}{f'}} \right\}\,>0\,.
\end{split}
\end{equation*}
Similar expressions are found for $\mu_{a}^{\pm}$ and $\mu_{b}^{\pm}$. We will write this briefly as
\begin{equation}
\lambda_{a}^{\pm}=  \frac{\Gamma_\pm(e,f)}{4|C(e,f)|}\,,\quad
\mu_{a}^{\pm}=  \frac{\Gamma_\mp(e,f)}{4|C(e,f)|}\,, 
\end{equation}
\begin{equation}
\lambda_{b}^{\pm}=  \frac{\Gamma_\pm(e',f')}{4(D(e,f)+\sqrt{|\Delta(e,f)|})}\,,\quad
\mu_{b}^{\pm}=  \frac{\Gamma_\mp(e',f')}{4(D(e,f)-\sqrt{|\Delta(e,f)|})}\,,
\end{equation}
\begin{eqnarray}
\Gamma_\pm(e,f)&\equiv& \mip{e}{f}\pm\sqrt{\mip{e}{e}\mip{f}{f}}>0\,,\\
\Gamma_\pm(e',f')&\equiv& \mip{e'}{f'}\pm\sqrt{\mip{e'}{e'}\mip{f'}{f'}}>0\,.
\end{eqnarray} 
It is now straightforward to observe given the slopes of the two straight lines intersecting the hyperbola
that the two segments $\mS_a,\mS_b$ are nonintersecting exactly when 
$\lambda_{a}^{-}\le\lambda_{b}^{+}$
and $\lambda_{b}^{-}\le\lambda_{a}^{+}$. But since the slope of the second line is always greater in magnitude
than that of the first line, the second of these inequalities is always satisfied (since the lines always intersect
the hyperbola). Thus we have:
\begin{lemma}
Let $e,f\in\mL$, with $e\sigma f$, $[e,f]\ne\nul$. Then
\begin{equation}
\mS_a\cap\mS_b\ne\emptyset\ \iff\ \lambda_{a}^{-}\le\lambda_{b}^{+}\,.
\end{equation}
\end{lemma}
\noindent This inequality is evaluated as follows:
\begin{equation}
\left(D(e,f)+\sqrt{|\Delta(e,f)|}\right)\,\Gamma_-(e,f)\le|C(e,f)|\,\Gamma_+(e',f')\,.\label{cond-a1}\\
\end{equation} 
We note the following relations: 
\begin{eqnarray}
\Gamma_+(e,f)\Gamma_-(e,f)&=&|C(e,f)|\,,\label{eqn:id1}\\
\Gamma_+(e',f')\Gamma_-(e',f')&=&|C(e',f')|\,,\label{eqn:id2}\\
\left(D(e,f)+\sqrt{|\Delta(e,f)|}\right)\left(D(e,f)-\sqrt{|\Delta(e,f)|}\right)&=&|C(e,f)C(e',f')|\,,\label{eqn:id3}\\
\Gamma_+(e,f)\,\Gamma_+(e',f')\Gamma_-(e,f)\,\Gamma_-(e',f')&=&|C(e,f)C(e',f')|\,.\label{eqn:id4}
\end{eqnarray}
Using these identities, (\ref{cond-a1}) is found to be  equivalent to
\begin{equation}
D(e,f)+\sqrt{|\Delta(e,f)|}\le\Gamma_+(e,f)\,\Gamma_+(e',f')\,.\label{cond-a2}\\
\end{equation}

We proceed to transform this inequality further. First rearrange terms so that only $\sqrt{|\Delta(e,f)|}$
remains on the left hand side, then square the expressions on both sides to obtain the inequality:
\begin{equation}\label{eqn:delgam1}
-\Delta(e,f)\le (\Gamma_+(e,f)\,\Gamma_+(e',f') -D(e,f))^2\,.
\end{equation}
Using the form of $\Delta(e,f)$ given in (\ref{Delta<0}), then eq.~(\ref{eqn:delgam1}) becomes 
after some rearrangement:
\begin{equation}
2D(e,f)\Gamma_+(e,f)\,\Gamma_+(e',f')\le(\Gamma_+(e,f)\,\Gamma_+(e',f'))^2+|C(e,f)C(e',f')|\,.\label{eqn:delgam2}
\end{equation}
Using (\ref{eqn:id4}), the last inequality becomes
after cancellation of $\Gamma_+(e,f)\,\Gamma_+(e',f')$:
\begin{equation}
2D(e,f)\le\Gamma_+(e,f)\,\Gamma_+(e',f')+\Gamma_-(e,f)\,\Gamma_-(e',f')\,.\label{eqn:delgam3}
\end{equation}
Now observe that this last inequality entails that $D(e,f)\le\Gamma_+(e,f)\Gamma_+(e',f')$ since otherwise
$D(e,f)>\Gamma_+(e,f)\Gamma_+(e',f')\ge\Gamma_-(e,f)\Gamma_-(e',f')$, in contradiction to (\ref{eqn:delgam3}).
Thus we can transform back to the equivalent (\ref{eqn:delgam1}), and using that 
$D(e,f)\le\Gamma_+(e,f)\Gamma_+(e',f')$, we finally obtain (\ref{cond-a2}).

We note that although this characterization of the coexistence of $e,f$ was deduced 
under the assumption $e\sigma f$, $[e,f]\ne\nul$, it is trivially fulfilled if these assumptions are
violated, since then the left-hand side of (\ref{eqn:delgam1}) is zero or negative.
Using the definitions of $D(e,f)$ and $\Gamma_\pm(e,f)$, $\Gamma_\pm(e',f')$,  inequality (\ref{eqn:delgam3})
can be given in explicit form, leading to the following result.
\begin{theorem}\label{thm:main}
Let $e,f\in\mL$. Then $e,f$ are coexistent if and only if
\begin{equation}\begin{split}
\mip{e}{e'}&\mip{f}{f'}-\mip{e}{f'}\mip{e'}{f}-\mip{e}{f}\mip{e'}{f'}\\
&=D(e,f)-\mip{e}{f}\mip{e'}{f'}\le\sqrt{\mip{e}{e}\mip{f}{f}\mip{e'}{e'}\mip{f'}{f'}}\,.\label{cond-a}
\end{split}
\end{equation}
\end{theorem}
We give yet another reformulation of the inequality (\ref{cond-a}) which highlights the the significance
of the noncommutativity of $e,f$. To this end we recall that $D(e,f)=[|\Delta(e,f)|+|C(e,f)C(e',f')|]^{1/2}$,
isolate this term in (\ref{cond-a}) on the left-hand side, and  square both sides of the inequality. This gives:
\begin{equation*}\begin{split}
|\Delta(e,f)|&\le\left(\mip{e}{f}\mip{e'}{f'}+\sqrt{NN'}\right)^2-|C(e,f)C(e',f')|\\
&\quad=\left(\mip{e}{f}\mip{e'}{f'}+\sqrt{NN'}\right)^2-\left(\mip{e}{f}^2-N\right)\left(\mip{e'}{f'}^2-N'\right)\\
&\quad=\left(\mip{e}{f}\sqrt{N'}+\mip{e'}{f'}\sqrt{N}\right)^2.
\end{split}
\end{equation*}
Here we have introduced the abbreviations 
\begin{equation}\label{NN'}
N\equiv\mip{e}{e}\mip{f}{f}\,,\quad N'\equiv\mip{e'}{e'}\mip{f'}{f'}\,.
\end{equation}
Thus we have established the following, recalling that $|\Delta(e,f)|=|\mip{d}{d}|\ve\times\vf|^2=
|\mip{d}{d}|\frac 14\|[e,f]\|^2$.
\begin{corollary}
Effects $e,f\in\mL\subset M_4$ are coexistent if and only if the following inequality holds:
\begin{equation}\label{cond-a-new}
-\tfrac 14\mip{d}{d}\,\|[e,f]\|^2\le\left(\mip{e}{f}\sqrt{\mip{e'}{e'}\mip{f'}{f'}}+
\mip{e'}{f'}\sqrt{\mip{e}{e}\mip{f}{f}}\right)^2\,.
\end{equation}
\end{corollary}
The left-hand side has been written in a way such that the inequality becomes automatically true in 
the trivial cases of coexistence where $e,f$ are not spacelike related. It is also manifest that the 
inequality holds if $e,f$ commute.

\section{A special case}

We consider a special case of interest where the two effects $e,f$ (as well as their complements have zero-components equal to $\frac 12$. This case was treated in \cite{Busch86}. 
\begin{corollary}
\begin{eqnarray}
&&e=\tfrac 12(\id+\mathbf{\tilde{e}}\cdot\mathbf{\sigma}),
\quad f=\tfrac 12(\id+\mathbf{\tilde f}\cdot\mathbf{\sigma})
\ \text{ are coexistent}\nonumber\\
&&\quad\Longleftrightarrow |\mathbf{\tilde e}|^2+|\mathbf{\tilde f}|^2\le1+(\mathbf{\tilde e}\cdot\mathbf{\tilde f})^2\label{coex3}\\
&&\quad\Longleftrightarrow  |\mathbf{\tilde e}\times\mathbf{\tilde f}|^2
\le (1-\mathbf{\tilde e}^2)(1-\mathbf{\tilde f}^2)\label{coex2}\\
&&\quad\Longleftrightarrow |\mathbf{\tilde e}+\mathbf{\tilde f}|+|\mathbf{\tilde e}-\mathbf{\tilde f}|\le 2\,.
\end{eqnarray}
\end{corollary}
\noindent{\em Proof.} 
For the above form of effects it is straightforward to verify that inequality (\ref{cond-a}) assumes the
explicit form (\ref{coex3}). \qed

The coexistence condition in the form (\ref{coex2}) has a simple operational meaning as explained in \cite{BuHei08}: the quantities $1-\mathbf{\tilde e}^2$ and $1-\mathbf{\tilde f}^2$
are measures of the {\em unsharpness} of $e,f$, so that according to this inequality the degrees of unsharpness of a coexistent pair of effects $e,f$ cannot simultaneously be made small if $e,f$ do not commute.

\section{Comparison with \cite{Liu-etal08} and \cite{StReHe08}}

In \cite{Liu-etal08} the coexistence of $e,f$ is expressed in the form of a single inequality 
which reads in our notation:
\begin{equation}\label{eqn:LL}
(1-F(e)^2-F(f)^2)\left(1-\frac{x^2}{F(e)^2}-\frac{y^2}{F(f)^2}\right)\le (xy-4\ve\cdot\vf)^2
\end{equation}
Here 
\begin{eqnarray*}
x&=&e_0-(1-e_0)=2e_0-1=\mip{e}{e}-\mip{e'}{e'}\,,\\
y&=&f_0-(1-f_0)=2f_0-1=\mip{f}{f}-\mip{f'}{f'}
\end{eqnarray*}
are measures of {\em bias} (for example, $x=0$ iff $e_0=1-e_0=\frac 12$), and
\begin{equation}
F(e)=\sqrt{\mip{e}{e}}+\sqrt{\mip{e'}{e'}}\,,\quad F(f)=\sqrt{\mip{f}{f}}+\sqrt{\mip{f'}{f'}}\,.
\end{equation}
We prove that $F(e)$ is a measure of the {\em unsharpness} of the effect $e$.
\begin{lemma}\label{lem:unsharpness}
Let $e\in\mL$. Then
\begin{equation}
0\le F(e)\le 1\,.
\end{equation}
Furthermore,
\begin{itemize}
\item[(a)] $F(e)=0$ iff $e_0=|\ve|=\frac 12$, that is, $e$ is a rank-1 projection (a nontrivial sharp effect);
\item[(b)] $F(e)=1$ iff $e=e_0\id$, that is, $e$ is a trivial effect.
\end{itemize}
\end{lemma}
\noindent{\em Proof.} 
Write
\begin{eqnarray}
F(e)^2&=&\mip{e}{e}+\mip{e'}{e'}+2\sqrt{\mip{e}{e}\mip{e'}{e'}}\nonumber\\
&=& 1+2\sqrt{\mip{e}{e}\mip{e'}{e'}} -\left[1-\mip{e}{e}-\mip{e'}{e'}\right]\nonumber\\
&=&1+2\sqrt{\mip{e}{e}\mip{e'}{e'}}-2\mip{e}{e'}\nonumber\\
&=&1+2\left[\sqrt{\mip{e}{e}\mip{e'}{e'}}-\mip{e}{e'}\right]\,.\label{eqn:Fe}
\end{eqnarray}
Now we consider $e,e'$ temporarily as elements of $M_3$ so that we  may write:
\[
\mip{e}{e}\mip{e'}{e'}-\mip{e}{e'}^2=\mip{e\times_oe'}{e\times_oe'}\le 0
\]
Note that $e,e'$ are timelike or lightlike and so this expression is zero if $e,e'$ are collinear; if $e,e'$ are
not collinear, then $e\times_o e'$ must be spacelike. (Recall that $\mip{e}{k}=0$ is only possible for timelike
or lightlike $e,k$ if both are lightlike and $k=\alpha e$ with $\alpha\ge 0$.)\\
Hence we have $\sqrt{\mip{e}{e}\mip{e'}{e'}}\le\mip{e}{e'}$, and therefore $F(e)\le 1$.

(a) $F(e)=0$ is equivalent to $e_0^2-|\ve|^2=0=(1-e_0)^2-|\ve|^2$, which holds iff $|\ve|=e_0=1-e_0=\frac 12$.

(b) From the above expression for $F(e)^2$ it is clear that $F(e)=1$ happens if and only if the term 
in square brackets is zero, that is, if and only if $e,e'$ are collinear, so that $e=e_0\id$.
\qed

In \cite{StReHe08} a measure $S(e)$ of the  {\em sharpness} of an effect $e$ is introduced that is 
crucial for the formulation of the coexistence condition. It is defined as follows (in our notation):
\begin{equation}
S(e)=2\left[\mip{e}{e'}-\sqrt{\mip{e}{e}\mip{e'}{e'}}\right]\,.
\end{equation}
In light of the calculation (\ref{eqn:Fe}) it is immediately seen that the sharpness $S(e)$ is closely related to $F(e)$:
\begin{equation}
S(e)=1-F(e)^2\,.
\end{equation}
The properties desired of a measure $S(e)$ of sharpness of an effect were proposed in a brief paper of one of the
present authors \cite{Busch07} and are satisfied in the present case as shown in \cite{StReHe08} and
evident from Lemma \ref{lem:unsharpness}:
\begin{eqnarray}
&&0\le S(e)\le 1\,;\\
&& S(e)=0\iff e \text{\ is a trivial effect;}\\
&& S(e)=1\iff e \text{\ is a nontrivial projection (nontrivial sharp effect);}\\
&& S(e)=S(e').
\end{eqnarray}

In \cite{Liu-etal08} it is shown that the condition (\ref{eqn:LL}) is equivalent to the condition found in
\cite{StReHe08}. We now proceed to establish the equivalence of (\ref{cond-a}) and (\ref{eqn:LL}). 
First we note:
\begin{equation*}
\frac{x^2}{F(e)^2} =\left(\sqrt{\mip{e}{e}}-\sqrt{\mip{e'}{e'}}\right)^2\,,\quad 
 \frac{y^2}{F(f)^2} =\left(\sqrt{\mip{f}{f}}-\sqrt{\mip{f'}{f'}}\right)^2\,.
\end{equation*}
A lengthy calculation gives the following reformulation of the left-hand side (abbreviated $LHS$)
of (\ref{eqn:LL}):
\begin{equation*}\begin{split}
LHS&=1+x^2+y^2+2(\mip{e}{e}+\mip{e'}{e'})(\mip{f}{f}+\mip{f'}{f'})-8\sqrt{NN'}\\
&=x^2+y^2-1+8\mip{e}{e'}\mip{f}{f'}-8\sqrt{NN'}\\
&=8\mip{e}{e'}\mip{f}{f'}-8\sqrt{NN'}+1+4(e_0^2+f_0^2)-4(e_0+f_0).
\end{split}
\end{equation*}
Next we note:
\[
xy-4\ve\cdot\vf=\mip{\id-2e}{\id-2f}=\mip{e-e'}{f-f'},
\]
so that the right-hand side ($RHS$) becomes:
\begin{equation*}\begin{split}
RHS&=\left(\mip{e}{f}+\mip{e'}{f'}-\mip{e}{f'}-\mip{e'}{f}\right)^2\\
&=8\mip{e}{f}\mip{e'}{f'}+8\mip{e}{f'}\mip{e'}{f}+1+4(e_0^2+f_0^2)-4(e_0+f_0).
\end{split}
\end{equation*}
Now it is immediately seen that $LHS\le RHS$ is equivalent to (\ref{cond-a}).

\section{Discussion}

We have deduced an inequality which constitutes a necessary and sufficient condition for the 
coexistence of a pair of qubit effects. Our formulation differs from the conditions obtained 
in \cite{StReHe08} and \cite{Liu-etal08}, which in turn had been shown to be equivalent
in the latter publication. The equivalence of the latter conditions with an earlier version of our result (given in arXiv:0802.4167v2 and reproduced here in the Appendix) had only been confirmed using 
numerical techniques \cite{Liu-etal08}. Here we have proven the equivalence analytically.

An important difference between our approach and the two other approaches lies in the 
fact that the latter are based the standard parametrization of the set of qubit effects, while 
in the present paper the the focus is on the geometric and order structures of the set of effects. 
This may be of use for the open problems of finding  coexistence conditions for more than two 
effects or two observables and for higher-dimensional Hilbert spaces as well as obtaining 
generic operational interpretations of such conditions.

\section*{Appendix: Alternative Formulation}

An alternative formulation of the main result that uses the basis vectors $g,g'$ was developed 
in an early version of the present work (arXiv:0802.4167). As the authors of \cite{StReHe08} 
and \cite{Liu-etal08} refer to this, it is reproduced here for comparison. We recall that in this 
formulation we are working under the assumption $e,f\in\mL$ with $e\sigma f$ and $[e,f]\ne \nul$.

We use  $g,g'$ to parametrize 
$a=\frac 12(e+f)-v=\frac 12(e+f)-\lambda g-\mu g'\in H$. 
Then the conditions (i)-(iii) of subsection \ref{subsec:coex-M3} read as follows. Eq.~(\ref{v-d}) is
automatically fulfilled since $g,\ g'$ are $\mip{}{}$-perpendicular to $d$ (eq.~(\ref{eqn:gpm-perp-d})).
The equation (\ref{hyp1}) for the hyperbola $\Ha$ becomes
\begin{equation}\label{hyp2}
H(\lambda,\mu)\equiv\lambda^2\mip{g}{g}+\mu^2\mip{g'}{g'}+2\lambda\mu\mip{g}{g'}+
\tfrac 14\mip{d}{d}=0.
\end{equation}
Note that inequality (\ref{Delta<0}), that is, $\Delta(e,f)<0$, is the determinant condition
ensuring that (\ref{hyp2}) describes a hyperbola in the $\lambda-\mu-$plane.

\begin{figure}
\includegraphics[width=\columnwidth]{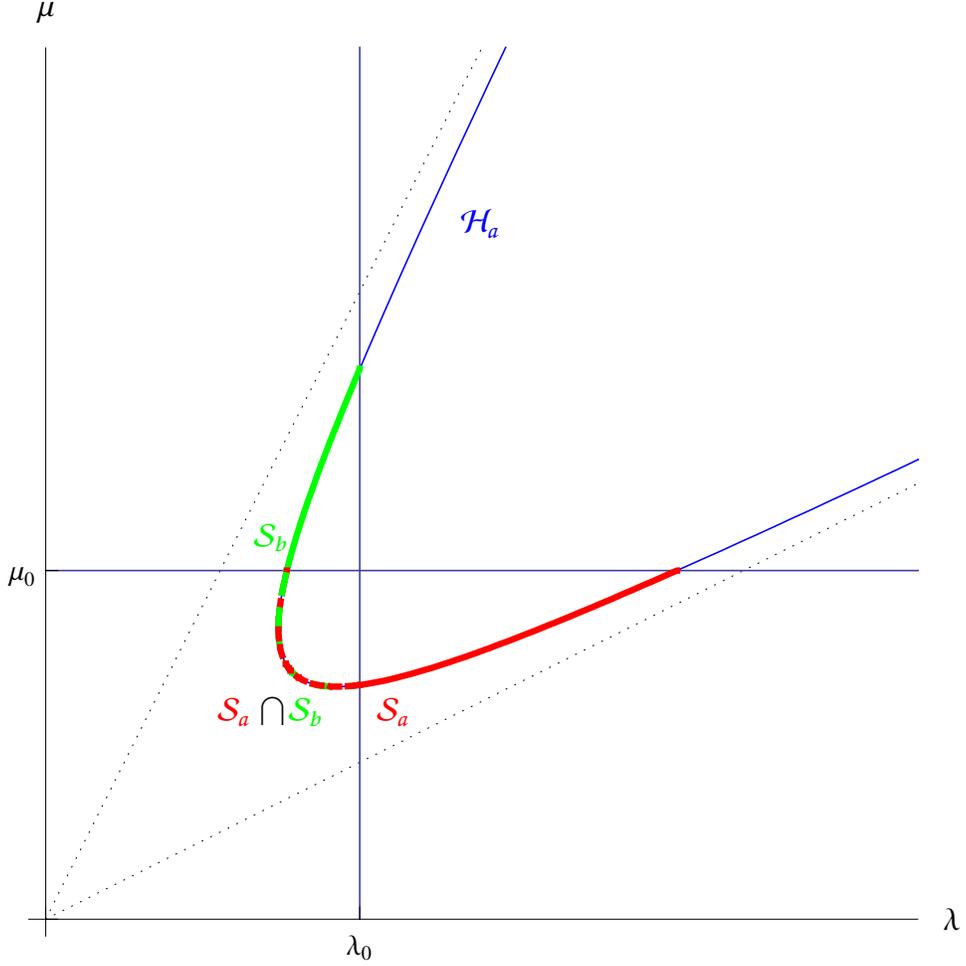}
\caption{\label{FIG_G2}One branch $\Hy_a$ of the hyperbola described by
Eq.(\ref{hyp2}) together with the segments $\mS_a$ and $\mS_b$
defined by $\mu\le\mu_0$ and $\lambda\le\lambda_0$, respectively.
To every point in $\mS_a\cap \mS_b$ there exist effects $a$ and
$b$ satisfying Eq.(\ref{coex-ef}) and hence $e$ and $f$ are coexistent.
}
\end{figure}

The conditions (\ref{vvdd1}) and (\ref{vvdd2}) for the end points of $\mS_a,\mS_b$ translate into
\begin{equation}\label{end-Sa}
\mu\mip{e+f}{g'}=\mip{e}{f}\quad\text{ for }\mS_a
\end{equation}
and
\begin{equation}\label{end-Sb}
\lambda\mip{e'+f'}{g}=\mip{e'}{f'} \quad\text{ for }\mS_b.
\end{equation}
Recalling eq.~(\ref{gpm}),
these linear equations now can be written in the form
\begin{eqnarray}
\mu&=&\frac{\mip{e}{f}}{2(e_1f_2-e_2f_1)}=\frac{\mip{e}{f}}{2|\ve\times\vf|} =
\frac 12\mip{e}{f}\sqrt{\frac{|\mip{d}{d}|}{|\Delta(e,f)|}}
\equiv\mu_0\quad\text{for }\mS_a\,,\\
\lambda&=&\frac{\mip{e'}{f'}}{2(e_1f_2-e_2f_1)}= \frac{\mip{e'}{f'}}{2|\ve\times\vf|}=
\frac12\mip{e'}{f'}\sqrt{\frac{|\mip{d}{d}|}{|\Delta(e,f)|}}
\equiv\lambda_0\quad\text{for }\mS_b\,.
\end{eqnarray}
These two equations describe a horizontal and a vertical line in the $\lambda-\mu-$plane each 
of which intersects the hyperbola (\ref{hyp2}), thus cutting out the two bounded segments $\mS_a$ 
and $\mS_b$, see Fig.~\ref{FIG_G2}.\footnote{We use the same notation $\Ha$, $\Ha_a$, $\mS_a$ and 
$\mS_b$ for the representations in the $\lambda-\mu-$plane of the hyperbola, its
branch $\B(e)\cap\B(f)$, and its segments  $\mS_a$, $\mS_b$.} 
By assuming $0<e_1f_2-e_2f_1=|\ve\times\vf|$, as we did in (\ref{gpm}), we have ensured  that $\Ha_a$
lies in the first quadrant ($\lambda\ge 0,\mu\ge 0$). 
We will eventually write the resulting inequalities in a symmetric
fashion with respect to $e$ and $f$; hence the case where $\Ha_a$ lies in the third
quadrant need not be considered separately.

\begin{figure}
\includegraphics[width=10cm]{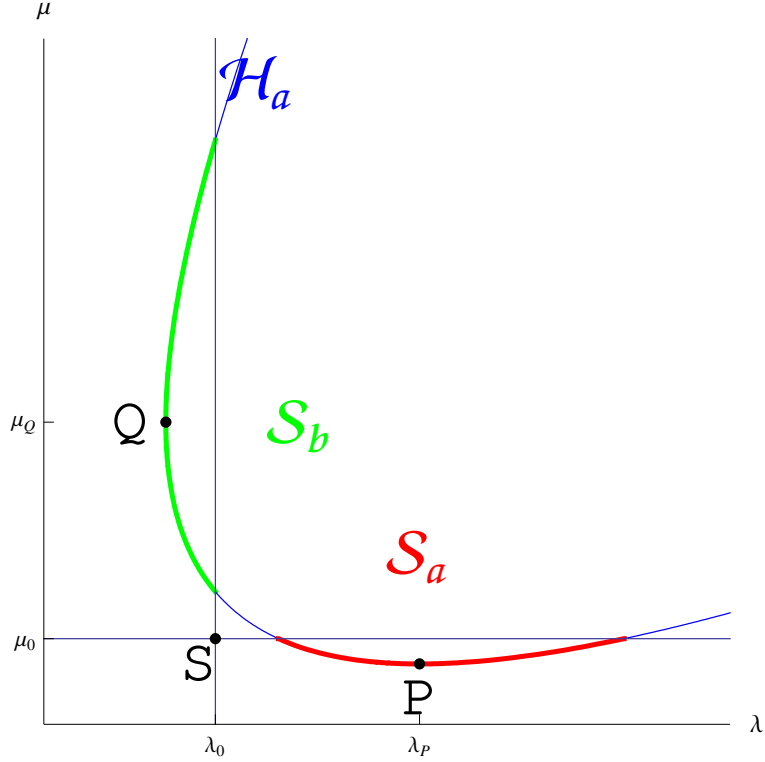}
\caption{\label{FIG_G3}The only case where $\mS_a\cap
\mS_b=\emptyset$ occurs if and only if $\lambda_0<\lambda_Q$ and $\mu_0<\mu_P$;
in this case $S=(\lambda_0,\mu_0)$ lies outside the convex hull of the
hyperbola branch $\Hy_a$.}
\end{figure}

The coexistence condition $\mS_a\cap\mS_b\ne\nul$ can be explored by considering
the location of the intersection point $(\lambda_0,\mu_0)$ of the two lines described by 
$\lambda=\lambda_0$ and $\mu=\mu_0$ relative to the hyperbola branch $\Ha_a$.
Let $P$ and $Q$ be the points of $\Ha_a$ which have a tangent parallel
to the $\mu$-axis and $\lambda$-axis, respectively,
with coordinates $(\lambda_P,\mu_P)$ and $(\lambda_Q,\mu_Q)$  (see Fig.~\ref{FIG_G3}).
After a short calculation we obtain
\begin{eqnarray*}
 \mu_P^2 &=& -\frac 14\frac{\mip{d}{d}\mip{g}{g'}^2}{\mip{g'}{g'}(\mip{g}{g}\mip{g'}{g'}-\mip{g}{g'}^2)}=-\frac{\mip{d}{d}D(e,f)^2}{4C(e',f')\Delta(e,f)},\\
 \lambda_Q^2 &= & -\frac 14\frac{\mip{d}{d}\mip{g}{g'}^2}{\mip{g}{g}(\mip{g}{g}\mip{g'}{g'}-\mip{g}{g'}^2)}=-\frac{\mip{d}{d}D(e,f)^2}{4C(e,f)\Delta(e,f)}.
\end{eqnarray*}

We have always:
\begin{equation}
P\in\mS_a \quad\text{and}\quad Q\in\mS_b.
\end{equation}
Hence,
\begin{eqnarray}
\mu_0\ge\mu_P&\Longrightarrow& \mS_a\cap\mS_b\ne\emptyset;\\
\lambda_0\ge\lambda_Q&\Longrightarrow& \mS_a\cap\mS_b\ne\emptyset.
\end{eqnarray}

In the remaining case of
\begin{equation}
\mu_0<\mu_P \quad\text{and}\quad \lambda_0<\lambda_Q
\end{equation}
we have (see Fig.~\ref{FIG_G3}):
\begin{equation}
\mS_a\cap\mS_b\ne\emptyset\
\Longleftrightarrow (\lambda_0,\mu_0)\in\mathrm{conv}(\Ha_a) 
\Longleftrightarrow
H(\lambda_0,\mu_0)\ge 0.\label{hyp-conv}
 \end{equation}

The inequalities 
$\mu_0^2\ge\mu_P^2$, $\lambda_0^2\ge\lambda_Q^2$ and $H(\lambda_0,\mu_0)\ge 0$ are 
collectively necessary and sufficient (though non-exclusive) conditions for the coexistence 
of $e,f$. With some rearrangement they assume the form:
\begin{eqnarray}
&&D(e,f)^2\ \le\  -\mip{e}{f}^2C(e',f')\,;\label{fin3a}\\
&&D(e,f)^2\ \le\ -\mip{e'}{f'}^2C(e,f)\,;\label{fin4a}\\
&&-\Delta(e,f)\le\ 
2\mip{e}{f}\mip{e'}{f'}D(e,f)
+\mip{e'}{f'}^2 C(e,f)+\mip{e}{f}^2 C(e',f')\, .
\end{eqnarray}
Noting (\ref{C<0}), (\ref{C'<0}), 
 the first two inequalities can also be written as:
\begin{eqnarray}
-\Delta(e,f)&\le&-\mip{e}{e}\mip{f}{f}\,C(e',f')\, ,\\
-\Delta(e,f)&\le& -\mip{e'}{e'}\mip{f'}{f'}C(e,f)\, .
\end{eqnarray}
These two inequalities are automatically satisfied if $e,f$ are not spacelike related or if they commute 
since in these cases $-\Delta(e,f)<0$. Hence we have established the following.

\begin{theorem}\label{thm:alt}
Let $e,f\in\mL\subset M_4$. Then $e$ and $f$ are coexistent if and only if at least one of the
following conditions is satisfied:
\begin{eqnarray}
&&-\Delta(e,f)\ \le\ \mip{e}{e}\mip{f}{f}\,|C(e',f')|\,;\label{final3}\\
&&-\Delta(e,f)\ \le\ \mip{e'}{e'}\mip{f'}{f'}|C(e,f)|\,;\label{final4}\\
&&-\Delta(e,f)\le\ 
2\mip{e}{f}\mip{e'}{f'}D(e,f)
-\mip{e'}{f'}^2 |C(e,f)|-\mip{e}{f}^2 |C(e',f')|\, .\label{final5'}
\end{eqnarray}
\end{theorem}

Finally we show explicitly that this formulation is equivalent to the main result.
We note first that inequalities (\ref{final3}), (\ref{final4}) and (\ref{final5'}) of 
Theorem \ref{thm:alt} can be rephrased in the equivalent form
\begin{eqnarray}
D(e,f)^2&\le& \mip{e}{f}^2|C(e',f')|\,;\label{final3'}\\
D(e,f)^2&\le& \mip{e'}{f'}^2|C(e,f)|\,;\label{final4'}\\
\big(D(e,f)-\mip{e}{f}\mip{e'}{f'}\big)^2&\le& NN'\,.\label{final5''}
\end{eqnarray}
Here we use the abbreviations (\ref{NN'}) for $N,N'$.
We also observe that the inequality (\ref{cond-a}) of Theorem \ref{thm:main} can be split up into two (non-exclusive) bits: (\ref{cond-a}) holds if and only if (\ref{final5''}) is true or
\begin{equation}\label{cond-aaa}
D(e,f)-\mip{e}{f}\mip{e'}{f'}\le 0\,.
\end{equation}

Assume the inequalities of Theorem \ref{thm:alt} hold. If the last one, (\ref{final5''}), holds, 
then (\ref{cond-a})  of Theorem \ref{thm:main} follows.

Next suppose that the last inequality (\ref{final5''}) is violated, so that one of (\ref{final3'}) or 
(\ref{final4'}) is valid. It follows readily each of them implies (\ref{cond-aaa}). Indeed, one of the
following two chains of inequalities applies:
\begin{eqnarray*}
D(e,f)\le\mip{e}{f}\sqrt{|C(e',f')|}<\mip{e}{f}\mip{e'}{f'}\,;\\
D(e,f)\le\mip{e'}{f'}\sqrt{|C(e,f)|}<\mip{e}{f}\mip{e'}{f'}\,.
\end{eqnarray*}

To prove the converse implication, we note that (\ref{cond-a}) is also equivalent to the  
exclusive alternative: either (\ref{final5''}), or
\begin{equation}\label{cond-aa}
D(e,f)-\mip{e}{f}\mip{e'}{f'}< -\sqrt{NN'}\,.
\end{equation}
We assume that (\ref{cond-aa}) holds. 
(In the other case the conclusion follows trivially.) Then (\ref{final5''}) is violated and so we have to
show that (\ref{final3'}) or (\ref{final4'}) follows. Suppose these are both violated, i.e.,
$D(e,f)^2>\mip{e}{f}^2|C(e',f')| $ and $D(e,f)^2>\mip{e'}{f'}^2|C(e,f)|$. Thus we obtain
\begin{eqnarray*}
\mip{e}{f}^2|C(e',f')|<D(e,f)^2< (\mip{e}{f}\mip{e'}{f'}-\sqrt{NN'})^2\,,\\
\mip{e'}{f'}^2|C(e,f)|<D(e,f)^2< (\mip{e}{f}\mip{e'}{f'}-\sqrt{NN'})^2\,.
\end{eqnarray*}
After some rearrangements and using 
\[
|C(e,f)|=\mip{e}{f}^2-N\,,\quad|C(e',f')|=\mip{e'}{f'}^2-N'\,,
\] 
this implies
\begin{eqnarray*}
2\sqrt{NN'}\mip{e}{f}\mip{e'}{f'}&<&NN'+N'\mip{e}{f}^2\,,\\
2\sqrt{NN'}\mip{e}{f}\mip{e'}{f'}&<&NN'+N\mip{e'}{f'}^2\,.
\end{eqnarray*}
Further rearrangement yields:
\begin{eqnarray*}
\sqrt{NN'}(\mip{e}{f}\mip{e'}{f'}-\sqrt{NN'})&<&\sqrt{N'}\mip{e}{f}[\sqrt{N'}\mip{e}{f}-\sqrt{N}\mip{e'}{f'}]\,,\\
\sqrt{NN'}(\mip{e}{f}\mip{e'}{f'}-\sqrt{NN'})&<&\sqrt{N}\mip{e'}{f'}[\sqrt{N}\mip{e'}{f'}-\sqrt{N'}\mip{e}{f}]\,,\\
\end{eqnarray*}
The expression on the left-hand sides is non-negative. The right-hand sides are not both non-negative,
so exactly one of them is non-positive, and the corresponding inequality is thus violated (even in the 
limiting case of $0<0$). This is a contradiction, and the proof is complete.

\end{document}